\pdfoutput=1
\documentclass{JINST}
\let\ifpdf\relax
\usepackage{url}
\usepackage{tabularx}

\newcommand{\figref}[1]{\figurename~\ref{#1}}
\newcommand{\secref}[1]{Section~\ref{#1}}

\newcommand{\equref}[1]{Equation~\ref{#1}}
\newcommand{\tabref}[1]{Table~\ref{#1}}
\usepackage{booktabs}
\usepackage{xspace}
\usepackage{pbox}
\usepackage{mathrsfs}
\usepackage{amsmath}
\usepackage[cal=rsfs,calscaled=.96]{mathalfa}

\newcommand{\dqdxz}{\ensuremath{\Delta Q_0/\Delta s_0}\xspace}

\newcommand{\dqdxo}{\ensuremath{\Delta Q_1/\Delta s_1}\xspace}

\title{Performance study of the effective gain of 
the double phase liquid Argon LEM Time Projection Chamber}

\author{C.~Cantini$^a$, L.~Epprecht$^a$, A.~Gendotti$^a$,
  S.~Horikawa$^a$, L.~Periale$^a$, S.~Murphy$^a$, G.~Natterer$^a$,
  C.~Regenfus$^a$, F.~Resnati$^a$, F.~Sergiampietri$^{a,b}$,
  A.~Rubbia$^a$\thanks{Corresponding
    author.}, T.~Viant$^a$ and S.~Wu$^a$~\\
  \llap{$^a$}ETH Zurich, Institute for Particle Physics,\\
  CH-8093 Z\"{u}rich, Switzerland\\
   \llap{$^b$}INFN-Sezione di Pisa\\
   56127 Pisa, Italy\\
  E-mail: \email{Andre.Rubbia@cern.ch}}

\abstract{The Large Electron Multipliers (LEMs) are key components of
  double phase liquid argon TPCs. The drifting charges after being
  extracted from the liquid are amplified in the LEM positioned half a
  centimeter above the liquid in pure argon vapor at 87 K. The LEM is
  characterised by the size of its dielectric rim around the holes,
  the thickness of the LEM insulator, the diameter of the holes as
  well as their geometrical layout. The impact of those design
  parameters on the amplification were checked by testing seven
  different LEMs with an active area of 10$\times$10 cm$^2$ in a
  double phase liquid argon TPC of 21 cm drift. We studied their
  response in terms of maximal reachable gain and impact on the
  collected charge uniformity as well as the long-term stability of
  the gain. We show that we could reach maximal gains of around 150
  which corresponds to a signal-to-noise ratio ($S/N$) of about 800
  for a minimal ionising particle (MIP) signal on 3 mm readout
  strips. We could also conclude that the dielectric surfaces in the
  vicinity of the LEM holes charge up with different time constants
  that depend on their design parameters.
  Our results demonstrate that the LAr LEM TPC is a robust
concept that is well-understood and well-suited for operation in ultra-pure cryogenic
environments and that can match the goals of
future large-scale liquid argon detectors.}


\begin{document}

\section{Introduction}\label{sec:introduction}
The double phase Liquid Argon Large Electron Multiplier Time Projection Chamber (LAr LEM TPC) 
is a 3D tracking and calorimetric device
which offers many advantages~\cite{Badertscher:2008rf, Badertscher:2009av,Badertscher:2011sy}. 
It has large density, very high granularity,
excellent energy resolution and ionisation signal amplification with tuneable gain. 
Unlike in the case of a single phase LAr TPC~\cite{Amerio:2004ze},
where the charge image is readout with wire planes in the liquid, the
double phase readout takes advantage of the charge multiplication in
gas argon for excellent signal-to-noise ratio on the single signal waveform, yielding optimal
image quality over large volumes and low energy threshold. The amplification of the drifting charges in gas allows
to build LAr TPCs to large scales and low threshold that are required by future long
baseline neutrino experiments. In this context, the Giant Liquid Argon
Charge Imaging ExpeRiment (GLACIER
~\cite{Rubbia:2004tz,Rubbia:2009md,Badertscher:2010sy}) is proposed as
far detector for the future European Long Baseline Neutrino Experiment
(LBNO~\cite{Stahl:2012exa}), enabling measurements 
over a wide range of energies from the
MeV to multi-GeV and addressing a broad spectrum of fundamental
physics topics, including long-baseline neutrino oscillations, 
proton decay search and supernova and atmospheric
neutrinos detection.
A large-scale demonstrator is being
currently planned in the context of the LBNO-DEMO (CERN WA105)
experiment~\cite{Agostino:2014qoa}.

\begin{figure}[htb]
  \centering
     \includegraphics[width=.8\textwidth]{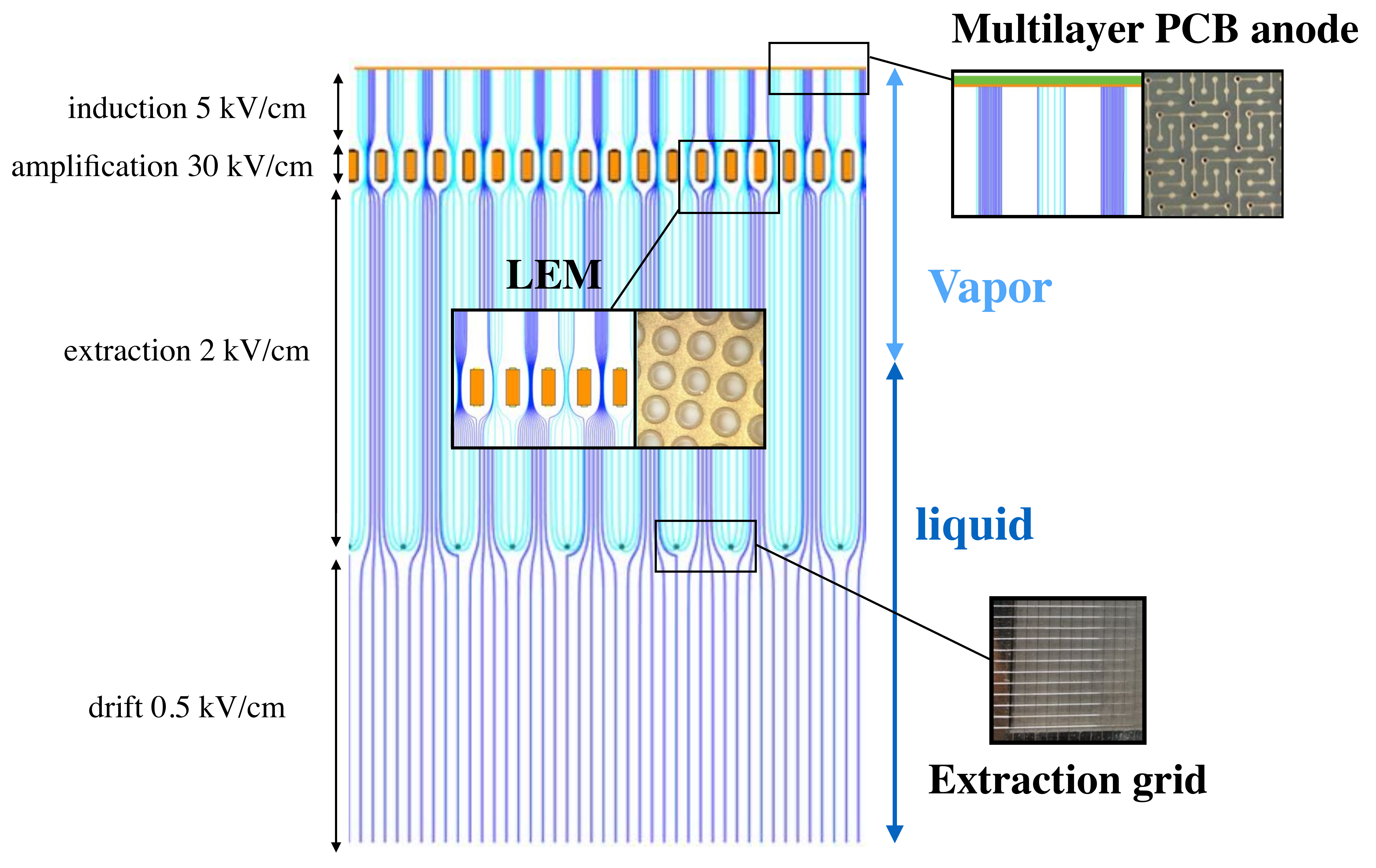}  
     \caption{Illustration of the amplification region in a double
       phase LAr TPC. The simulated field lines in dark blue are an
       indication of those followed by the drifting charges (without
       diffusion).}
     \label{fig:larltemtpc} 
   \end{figure}

In the double phase LAr LEM TPC, the ionisation charge is extracted
to the Argon vapour, where it is amplified by a LEM stage, which triggers Townsend multiplication in the high
electric field regions in the LEM holes ~\cite{Bondar:2008yw}. See Figure~\ref{fig:larltemtpc}.
After their long drift
within the liquid phase with a drift field of typically 0.5~kV/cm, ionisation
electrons are efficiently extracted from the liquid with an electric
field of around 2 kV/cm and amplified with a field of about 30 kV/cm
applied across both electrodes of the LEM. The amplified charge is
then collected and recorded on a two-dimensional and segmented anode.
The anode consists of a set of strips (views) that provide the $x$ and
$y$ coordinate of the event with a mm-level~granularity.  

The double phase principle was repeatedly successfully demonstrated on a chamber
equipped with a $10\times10$ cm$^2$ area readout (see e.g
Refs. \cite{Badertscher:2008rf,Badertscher:2010fi} ) and on a larger
device consisting of a $40\times80$~cm$^2$ readout and 60 cm drift
\cite{Badertscher:2013wm}. We were able to operate both setups in a
stable condition at constant gains of about 15 corresponding to
$S/N\approx 60$ for MIPs. The possibility of
obtaining such large signal-to-noise ratios is very appealing,
considering the fact that increasing detector sizes with longer drifts
and larger readout capacitances lead to a degradation of the imaging
quality of the device. Future large double phase TPCs will be composed
of anode and LEM modules of 50$\times$50 cm$^2$ area. In
Ref. \cite{Cantini:2013yba} we demonstrated that for large areas the
electronic noise could be kept within $\sim$1000 electrons for a two
meter readout by using an innovative design of multilayer PCB anodes.

In this paper we study in more detail the effective gain and
investigate in which manner the design specifications of the LEM
affect the amplification of the drifting charges. The relevant
features of the LEM are 1) the diameter of the hole, 2) the thickness
of the insulator, 3) the size of the dielectric rim around
the holes and 4) the geometrical arrangement of the holes. The LEM
parameters potentially affect the effective gain of the chamber either
by changing the electric field configuration inside the LEM hole or
modifying the electrical transparency of the system.

\section{Experimental Setup}
\begin{figure}[htb]
  \centering
     \includegraphics[width=\textwidth]{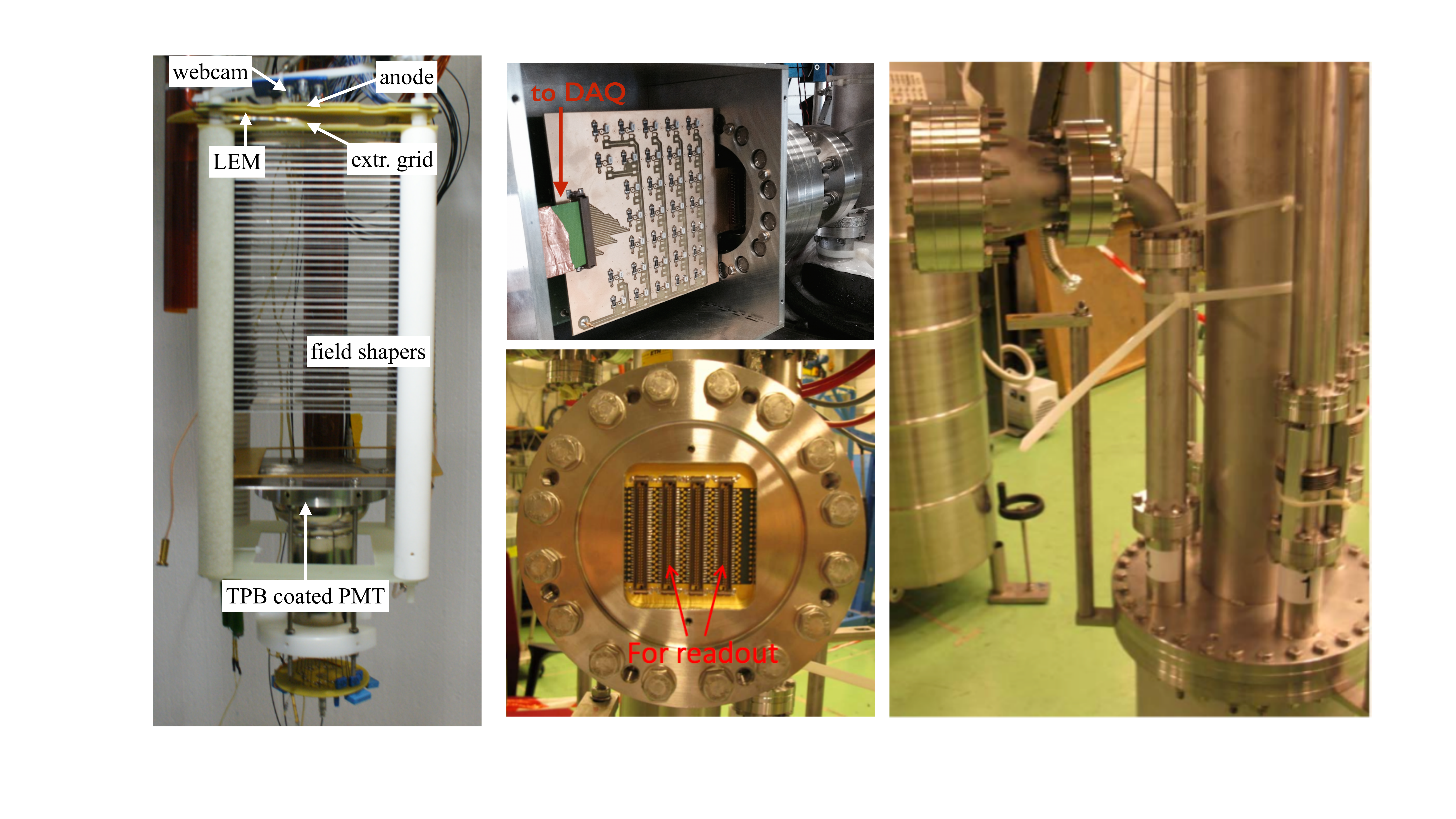}  
     \caption{Picture of the 3 liter TPC (left) and of the stainless
       steel chimney through which the signal cables are brought
       outside of the cryostat to the decoupling boards and to the
       DAQ.}
     \label{fig:setup} 
   \end{figure}

The LEMs are tested in the the so-called ``3 liter'' setup which is a
double phase LAr LEM TPC consisting of a 21 cm long drift length and a
$10\times10$ cm$^2$ area. The setup has now been operated for more
than 4 years, it is a well understood detector and very useful for
testing new ideas with a rapid turn-around. A description of the
apparatus can be found in Ref.~\cite{Cantini:2013yba}. In
\figref{fig:setup} we show a picture of the drift cage along with
photos detailing the charge readout system.

   The free charges from the ionisation are drifted towards the LAr
   surface driven by a constant drift field of 500 V/cm. A larger
   electric field of about 2 kV/cm is applied at the vicinity of the
   LAr surface to efficiently extract the drifting charges to the
   vapour phase. The extraction field is confined in a 10~mm region
   between the LEM and a stainless steal grid placed in the liquid.
   The grid consists of 100 $\mu$m stainless steel wires tensed and
   positioned at a constant pitch of 3 mm in both $x$ and $y$
   direction matching the anode strips. This configuration guarantees
   that close to 100 \% of the drifting charges are extracted and
   focussed towards the LEM holes. The amplification and readout of
   the charge is then performed by the LEM and anode.  An electric
   field of at least 30~kV/cm is applied across the LEM in order to
   obtain charge amplification in the dense pure argon vapour. The electrons are collected on the
   anode placed 2~mm above the LEM. 
   
   In view of larger scale operation
   we tested a configuration whereby the high voltage decoupling of
   the readout and discharge protection system are all mounted on two
   dedicated PCBs placed immediately outside of the chamber. In this
   configuration the 64 signal cables are routed out via two twisted
   pair cables of 32 channels through a stainless steel chimney that
   is terminated with an ultra-high vacuum leak-tight
   feedthrough. Pictures are shown in \figref{fig:setup}-right. The
   flange is actually made of a Stainless Steel ring, in which a
   several mm-thick multilayer PCB is sealed. The multilayer PCB is
   designed in order to accommodate connectors on both faces. We
   believe this system to be scalable to flanges of larger diameters
   that can accomodate a higher number of channels as developed 
   in Ref.~\cite{Agostino:2014qoa}. The signals are then amplified and digitised by the
   specially developed CAEN SY2791 readout system (see
   Ref.~\cite{Badertscher:2013wm}).

   Before operation the liquid level is precisely adjusted between the
   extraction grid and the LEM by monitoring the capacitance between
   both systems. During filling for the first time were able to
   visually inspect the rise of the liquid argon level by installing a
   webcam combined with LED lighting above the anode. The webcam is a
   Microsoft VX-1000 described in Ref.~\cite{Mavrokoridis:2014gka}. It
   produces a clear image of the level during the filling. However
   once the chamber is filled and the camera reaches the temperature
   of argon vapour, the image is no longer visible. The camera becomes
   operational again once at room temperature. For future operations
   of large tanks we envisage to embed the camera in a dedicated
   casing with heaters to ensure a constant visual monitoring.

   In order to keep the chamber at liquid argon temperature, it is
   fully immersed in an open bath filled with liquid argon. In these
   conditions the chamber is operated at a stable pressure of around
   980 mbar slightly above the atmospheric pressure in our lab of 960
   mbar. The detector is however subject to slight pressure variations
   resulting from external modifications from the weather conditions. In addition the
   pressure rises periodically when the level of the open bath
   decreases below the top flange of the detector before it is
   re-filled by an operator. These pressure variations affect the operations
   of the chamber and will be corrected for as described later.

\subsection{The Large Electron Multipliers}\label{sec:lem_intro}
\begin{figure}[htb]
  \centering
     \includegraphics[width=.9\textwidth]{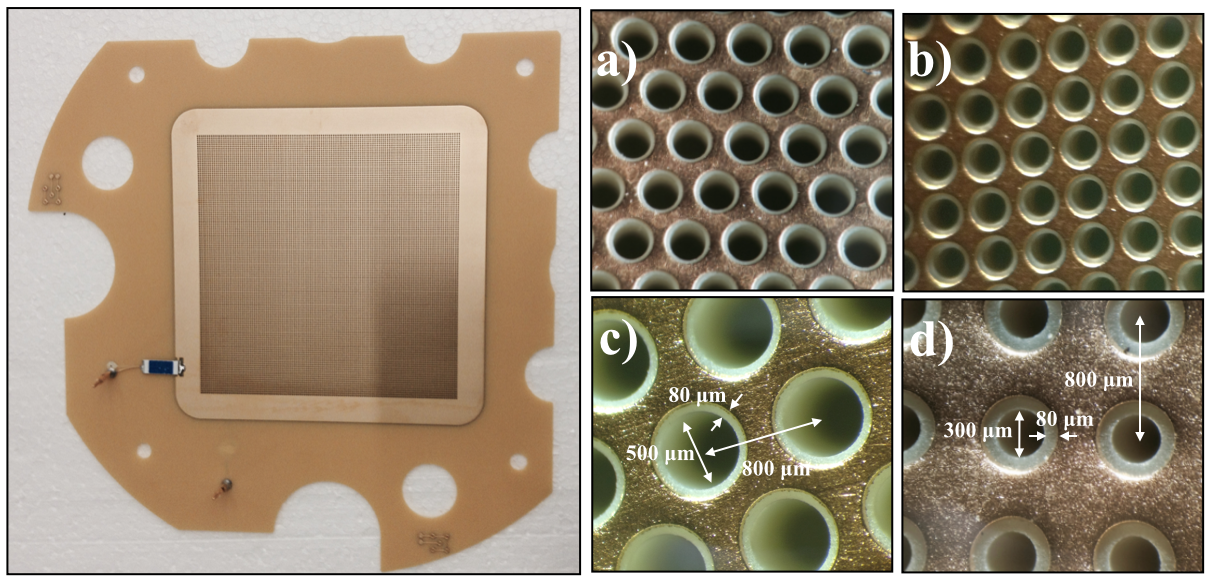}  
     \caption{Left: picture of one of the 10$\times$10 cm$^2$ LEMs we tested in
       our chamber. A microscope view of some LEM samples is shown on
       the right: the hexagonal hole arrangement (\textit{a}) is
       compared to the square arrangement (\textit{b}) and on the
       bottom zooms (\textit{c}) and (\textit{d}) on two samples with different hole sizes are
       shown.}
     \label{fig:lem_pics} 
\end{figure}

The Large Electron Multipliers (LEM) consist of copper cladded epoxy
plates with a thickness of a millimeter and with mechanically
drilled holes. The holes have diameters of 500~$\mu$m and there are order of 200 holes per cm$^2$.  
Applying a sufficient potential difference to the two metal faces of about 30~kV/cm, an
electric field strong enough to trigger the Townsend avalanche
is attained in the holes. The photons produced during the avalanche
are absorbed by the dielectric walls of the holes, ensuring the
confinement of the multiplication in absence of quenching gases. The
LEM is also well suited for the operation in cryogenic conditions
because the thermal expansion coefficient of the metal and of the
glass epoxy are well matched. It sustains without any damage abrupt
cooling down and warming up cycles, and is a device known to be
resistant to discharges. The 10$\times$10 cm$^2$ LEMs used for our
measurement are produced at a PCB manufacturing company called
ELTOS\footnote{\url{www.eltos.it}}. Pictures of some samples are shown
in \figref{fig:lem_pics}.  About 15'000 holes are mechanically drilled
with standard PCB techniques in the copper cladded glass epoxy
plate. In order to reduce the probability of discharge at the edge of the holes,
a dielectric rim is produced by etching away the copper
around the periphery of these latter. In order to guarantee uniform
sizes and perfect entering, the rims are produced by a mask-less etching technique
developed at CERN \cite{Breskin2009107}. The rims are typically a few
tens of micrometer large, their size is limited by the initial
thickness of copper. The metallisation of the LEM extends about one
centimeter around the area with the holes in order to properly shape
the electric field at the edge of the chamber active area.

The parameters that characterise the LEMs are the hole diameter, the
thickness of the epoxy plate, the geometrical arrangement of the holes
and the size of the rim. The complete list of the LEMs we tested along
with their individual specifications are provided in
\tabref{tab:lem_spec}. The precision on the parameters are those
quoted by the company and verified independently by us on some
samples.

\begin{table}[htb]
\renewcommand{\arraystretch}{1.3}
\renewcommand{\tabcolsep}{2mm}
\begin{center}
\begin{tabular}[\textwidth]{p{2.75cm}lccccccc}
  \toprule 
  &precision& LEM1 & LEM2 &LEM3 & LEM4 &  LEM5 & LEM6 & LEM7\\
  \midrule  
  rim size ($\mu$m)&$\pm$3 &80&80&40&80&40&80&80\\
  FR4 thickness (mm)&$_{-0.06}^{+0}$ &0.8&1&1&1&1&1&0.6\\
  copper thickness ($\mu$m)&$\pm 5$&80&80&40&80&40&80&80\\
  hole diameter ($\mu$m)&$_{-25}^{+0}$&500&500&500&400&500&300&500\\
  hole layout
  ($\mu$m)&-&square&square&hexagonal&square&square&square&square\\
  number of holes&-&15084&15084&16761&15084&15084&15084&15084\\
  $\kappa$ (see text)&-&0.789&0.830&0.900&0.905&0.900&0.938&0.699\\
  measured capacitance (pF)&$\pm$10&539&450&479&500&531&557&560\\
  \bottomrule 
\end{tabular}
\end{center}
\caption{\label{tab:lem_spec} Specifications of the LEMs tested
  in our setup.}
\end{table}

\begin{figure}[htb]
  \centering
     \includegraphics[width=\textwidth]{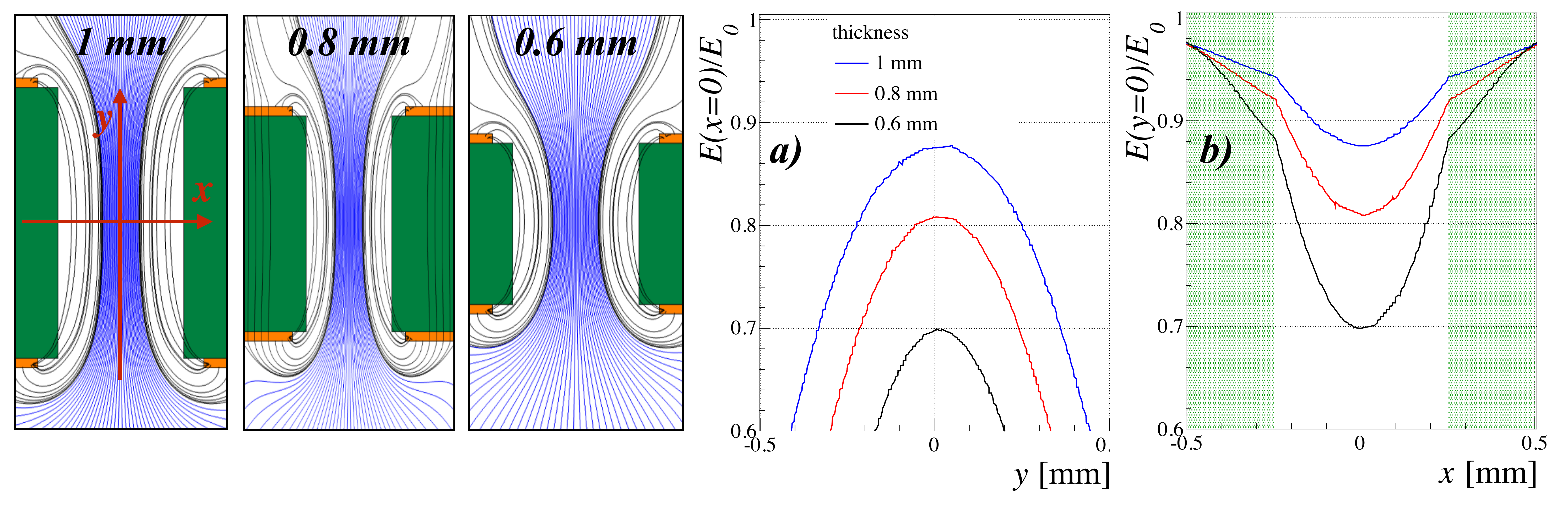}  
     \caption{Computation of the electric field lines inside a LEM
       hole for three different thickness of the insulator indicated
       on the figure (all the LEMs have a rim 80 $\mu$m and a hole
       diameter of 500 $\mu$m). The field lines followed by the
       drifting charges are shown in blue. The two graphs, $a)$ and
       $b)$, show the ratio of the norm of the electric field at the
       center of the hole ($x=0$, resp. $y=0$) to the field applied
       across the electrodes ($E_{0}$) as a function of the $y$
       (resp. $x$) coordinates. The shaded green area delimites the
       hole edges.}
     \label{fig:lem_comsol} 
   \end{figure}

   The multiplication of the electrons in the LEM holes depends on the
electric field created by applying a voltage $V$ across the
LEM of thickness $d$.  We present our results as a function of the 
effective electric field $E_0=V/d$, in reality 
the field inside the hole is always weaker than the nominal $E_0$.
   The computation of the field along the hole axis shown in graph
   $a)$ of \figref{fig:lem_comsol} indicates that the field is larger at the center of the
   hole. However, as shown in graph $b)$, the field is even greater on
   the side of the holes close to the insulator. The charge multiplied
   in the center region of the hole is efficiently transferred to the
   anode, while the one produced in the side region is more prone to
   be collected on the top electrode. Therefore the larger field on
   the hole edges may be responsible for discharges without
   significantly contributing to the amplification. 
   
We define $\kappa$
   as the ratio of the field at the center of the hole ($x=0 ,y=0$) to
   the applied field $E_0$ and $E\equiv\kappa E_0$ as the
   amplification field.  The effective gain $G_{eff}$ of the LEM is
   then generally expressed with the
   function~\cite{Badertscher:2011sy}:
\begin{equation}\label{eq:gain_func}
G_{eff}(E,\rho,t) \equiv  
{\cal T} e^{\alpha(\rho,E) x}\times {\cal C}(t)
\end{equation}
where ${\cal T}$ is a parameter proportional to the electrical
transparency of the chamber; $\alpha(\rho, E)$ is the first Townsend
ionisation coefficient for the amplification field $E$ and density
$\rho$; $x$ denotes the effective amplification length which can be
geometrically related to the length of the field plateau along the
hole axis shown in graph $b)$ of \figref{fig:lem_comsol}; and ${\cal
  C}(t)$ represents any time variation of the gain.  The generalised
form of the first Townsend coefficient as a function of the medium
density $\rho$ and the electric field $E$ can be approximated by
\cite{Aoyama:1985}:
\begin{equation}
\alpha(\rho, E)=A \rho  e^{-B \rho/E}
\end{equation}
where $A$ and $B$ are parameters depending on the gas.  A fit to the
electric field dependence of the Townsend coefficient in the range
between 20 and 100~kV/cm predicted by MAGBOLTZ~\cite{magboltz}
calculations, gives $A \rho =(7339\pm 90)$ cm$^{-1}$ and $B \rho
=(183\pm 1.0)~$kV/cm for pure argon at 87~K and 0.980~bar. We
refer the reader to Ref.~\cite{filippo_thesis} for a 
detailed explanation of charge amplification inside the LEM hole. 

The results from electrostatic
simulations presented in \figref{fig:lem_comsol} indicate that both
the amplification field and length increase with thicker LEMs. The
same simulations show that they also increase with smaller rim sizes
and smaller hole diameters. However those simulations do not take into
account parameters that affect the effective gain such as, for
instance, the electrical transparencies and breakdown voltages of the
LEMs. The results from the data presented in the next sections fully
characterises the LEMs in terms of their maximal effective gain as
well as time evolution of the effective gain in pure Argon at 87 K.

\subsection{Measurement of the effective gain}

The 3L chamber was exposed to large samples of cosmic muons tracks which were used to
characterise the response of the chamber in terms of effective gain,
signal-to-noise ratio and energy resolution. 
The analysis (noise filtering, hit-finding, track reconstruction
  etc..) of all the data  is performed with the QScan software package
  \cite{Rico:2002,devis_thesis}. The reconstruction method is similar to that
  described in Ref.~\cite{Badertscher:2013wm} and includes also an algorithm that searches
  for residual hits which are interpreted as $\delta$-rays a of few MeV
  emitted by the traversing ionising particles.
The crossing muons, once
reconstructed in 3D, allow us to retrieve the
length of the track on each strip of view 0 and view 1 ($\Delta s_0$
and $\Delta s_1$), along with the charge collected on the
corresponding channels (corrected for the electron lifetime), $\Delta
Q_0$ and $\Delta Q_1$ . The charge collected by unit length \dqdxz and
\dqdxo, which are proportional to the energy locally deposited by the
track in liquid Argon, are the relevant quantities used to evaluate
the performance of the LEM and estimate the gain of the chamber. Since
the cosmic muons that cross the chamber are minimum ionising particles
the average charge deposition along a track, predicted by the
Bethe-Bloch formula and accounting for electron-ion
recombination~\cite{Amoruso:2004dy} is $\langle \Delta Q/\Delta
s\rangle_{MIP} =10$ fC/cm. By using the sum of the collected charge
per unit length on both views we hence define the measured effective
gain by:
\begin{equation}
G_{eff}=\frac{\langle \dqdxz \rangle +\langle \dqdxo\rangle}{\langle
  \Delta Q/\Delta s\rangle_{MIP}}\label{eq:gain_meas}
\end{equation}
$G_{eff}$ takes into account the charge multiplication in the LEM
holes, as well as potential charge reduction from the liquid-vapour
extraction efficiency and from the transparency of the grid and the
LEM.

\section{Influence of various LEM parameters on the effective gain}
\subsection{Experimental procedure}
Before installation the LEM is washed with high-pressure deionised
water and further cleaned in an ultrasonic bath filled with pure
alcohol. In order to test the absence of resistive contacts between
two electrodes, the LEM is powered in air. Typical fields of about
35-40 kV/cm should be reached and the measured
leakage current should not exceed 10 nA.
The voltage is then increased further until discharges occur, their
location are monitored to be sure that there is no particular region
more susceptible to spark.  The controlled sparks train the LEM since
they remove dust and burn the fibers inside the hole.

The LEM is then mounted on the chamber and the high voltage is
supplied to the two electrodes through metal pins (see
\figref{fig:lem_pics}). The 500 M$\Omega$ resistor between the high
voltage connector and the electrode acts a spark protection device by
limiting the current across the hole during the discharge. Before
filing with liquid argon, the main vessel containing the chamber
is first evacuated to residual pressure below 5$\times$10$^{-6}$ mbar
and filled with pure liquid Argon trough an activated copper and
zeolite powder cartridge. The level of the liquid is precisely
adjusted in-between the extraction grid and LEM by monitoring the
capacitance of the LEM-grid system. The webcam provides complimentary
visual check of the liquid level. The liquid Argon purity is monitored
throughout the data taking period by measuring the lifetime of the
drifting electrons. The purity is typically below 1 part ber billion
(ppb) after filling and decreases at a rate of about 1.4 ppb per day
mainly due to the outgassing of material in the relatively small gas
volume of the chamber.

The seven LEMs listed in \tabref{tab:lem_spec} were tested in seven
independent runs with the same experimental procedure. We first
performed electric field scans of the LEM: at each field setting across
the LEM cosmic data was acquired for $\sim$ 15 minutes at a rate of
about 5 Hz. The field was ramped up by steps of 1 kV/cm until the
breakdown voltage was reached. For each LEM setting the induction and
extraction fields were always adjusted at the nominal values of 5 kV/cm
and 2 kV/cm in the liquid respectively. The electric field scan was the
first data-set to be recorded after the chamber is filled, meaning
that the LEM was polarised for the first time during the scan. Once the
scan was finished we acquired data continuously for a period of a few
days at constant electric field settings. The chamber was then emptied,
opened to air and the next LEM was mounted. The only differences
between the runs were the LEMs and the fields at which they were
operated. In the next two sub-sections the results of the electric
scans and the longer-term operation are discussed.

\subsection{Gain as a function of the applied electric field}\label{sec:lem_scan}
\begin{figure}[h!]
     \centering
     \includegraphics[width=\textwidth]{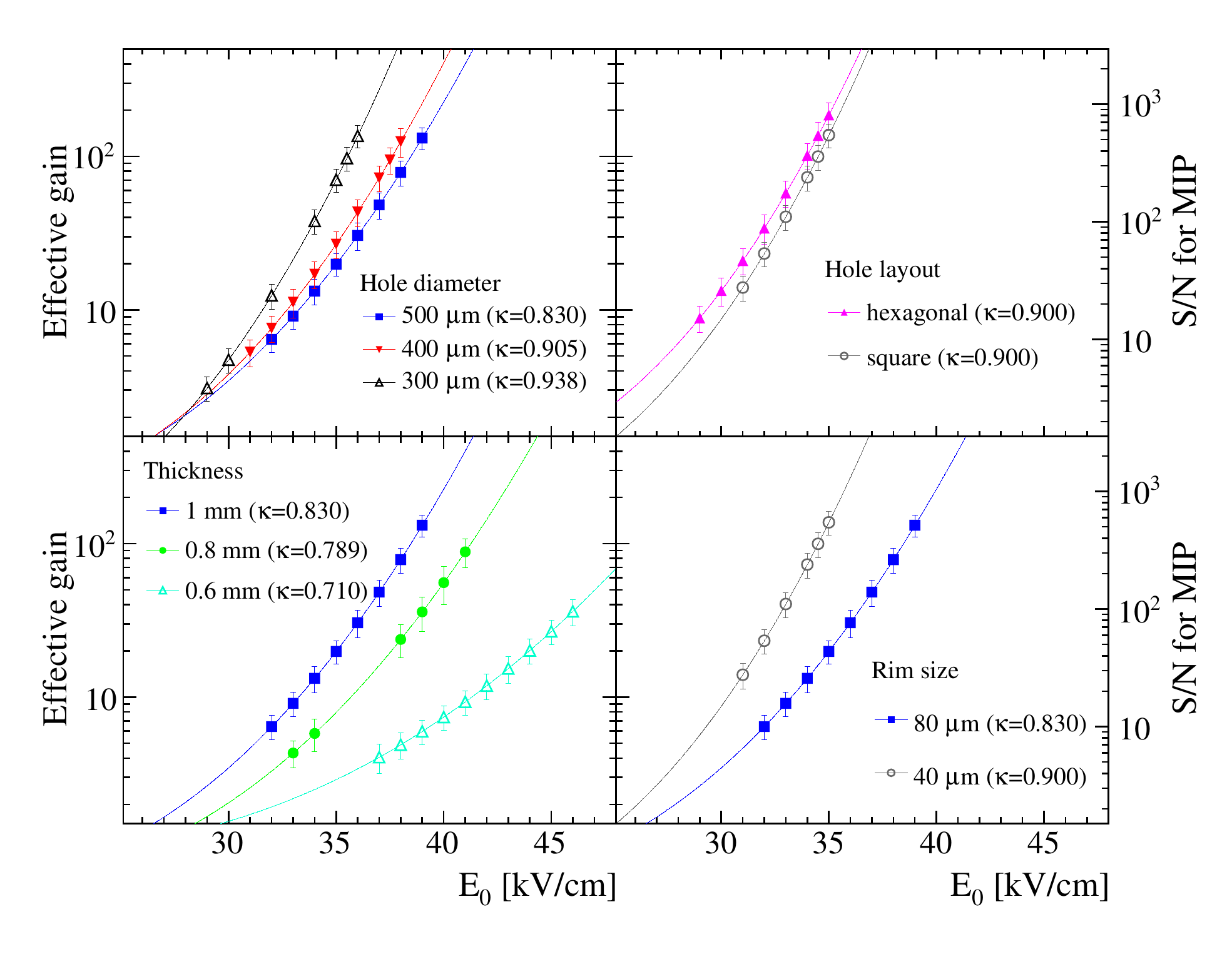} 
     \caption{Comparison of effective gains for various LEMs in pure
       argon vapor at 87 K as a function of the electric field applied
       across the electrodes. The data points are fitted with the
       function described in the text.}
     \label{fig:lem_scan_comp}
   \end{figure}

In this section we compare the results of electric field scans
obtained on all the LEMs.  The results are presented in
\figref{fig:lem_scan_comp} where the effective gain is plotted as a
function of the applied electric field across the LEM electrodes
($E_0$). The results are grouped together in such a way that the only
difference between the LEMs is the parameter indicated in the legend.
   The effective gain is corrected for the electron lifetime and
   normalised to a pressure of 980 mbar. The data points are fitted
   with the function described in \equref{eq:gain_func}. Since the LEM
   field scans occur on a short timescale of typically 2 hours we can
   neglect time variations of the gain ${\cal C}(t)$ during this
   period. This point will be justified in the next section. As can be
   seen, for all the LEMs the measured effective gains exhibit the
   expected exponential dependence with the LEM field. During the fit
   $\kappa$ was fixed at the values from the electrostatic field
   simulation results and $\cal{T}$ together with the amplification
   length, $x$, were left as free parameters. For all measurements the
   maximum reachable gain was always limited by the occurrence of
   discharges. The highest electric field setting at which data could
   be stably recorded is defined as $E_0^{max}$ and the corresponding
   gain as $G_{eff}^{max}$.  The results are summarised in
   \tabref{tab:lem_scan_summary} and are discussed below:

\begin{table}[htb]
\renewcommand{\arraystretch}{1.3}
\begin{center}
\begin{tabular}[\textwidth]{llllllc}
  \toprule
  tested parameter& value &LEM &\phantom{abcd}${\cal T}$ &$x$ (mm)&
  $G_{eff}^{max}$ & $E_0^{max}$ (kV/cm)\\
  \midrule
  hole layout&hexagonal&3&0.59$\pm$ 0.18&0.96$\pm$0.07&182&35\\
  &square&5&0.34$\pm$ 0.14&0.94$\pm$0.08&123&35\\
  \midrule
  hole diameter&500 $\mu$m &2&0.46$\pm$ 0.14&0.73$\pm$0.05&124&39\\
  &400 $\mu$m &4&0.41$\pm$ 0.11&0.81$\pm$0.05&124&38\\
  &300 $\mu$m &6&0.20$\pm$ 0.03&0.88$\pm$0.04&134&36\\
  \midrule
  thickness&1 mm   &2&0.46$\pm$ 0.14&0.73$\pm$0.05&124 &39\\
  &0.8 mm&1&0.46$\pm$ 0.15&0.69$\pm$0.06&88&41\\
  &0.6 mm&7&0.58$\pm$ 0.2&0.55$\pm$0.06&36&46\\
  \midrule
  rim size &40 $\mu$m &5 &0.34$\pm$ 0.14&0.94$\pm$0.08&123&35\\
  &80 $\mu$m &2&0.46$\pm$ 0.14&0.73$\pm$0.05&124&39\\
  \bottomrule
       \end{tabular}
     \end{center}
     \caption{\label{tab:lem_scan_summary} summary of the results retrieved from the
       LEM electric field scans. The definitions of the variables are
       given in the text.}
   \end{table}
   \begin{itemize}

   \item[(i)] {\it Impact of the hole geometrical layout}: the
     electrical transparency is significantly higher for the LEM with
     the hexagonal hole layout. They also have near identical
     amplification lengths. This can be understood graphically from
     the fact that both curves remain parallel. Since both LEMs
     breakdown at the same electric field the one with the hexagonal
     hole layout has a significantly larger $G_{eff}^{max}$ of almost
     200.

   \item[(ii)] {\it Impact of the hole diameter}: electrostatic
     simulations predict a slightly larger amplification length as the
     hole sizes decreases. We indeed observe this phenomena which
     translates into a faster rise of the effective gain with the
     electric field. We also notice that, due to the increased
     amplification field, at a given $E_0$ the gain is larger for the
     LEMs with smaller holes. However the results from the fit also
     indicate that the electrical transparency of the LEM reduces with
     smaller holes. We do not observe any improvement of the LEM
     performance in terms of maximal gain. All three LEMs are able to
     reach gains of about 120.

   \item[(iii)] {\it Impact of the thickness:} we see a clear enhancement
     of the slope for thicker LEMs due to the increase of the
     amplification lengths. We also observe that the maximal reachable
     gains are significantly lower for thinner LEMs.

   \item[(iv)] {\it Impact of the rim size:} we qualitatively observe
     an increase of the amplification length with the LEM of smaller
     rim size. However although not shown on the Figure, we observed
     from previous data that the maximal reachable gain was typically
     a factor two lower if we do not etch a rim around the LEM
     holes. The importance of the rim although in other gases and
     temperatures is discussed e.g in
     Ref. \cite{Breskin2009107}. Based on these results we do not see
     a benefit of going to a larger than 40 $\mu$m rims. On the
     contrary since for larger rims the amplification field is lower
     we need to apply higher voltages acrross the LEM to reach
     comparable gains.
\end{itemize}

\subsection{Long term stability and charging up}
\begin{figure}[h!]
     \centering
     \includegraphics[width=\textwidth]{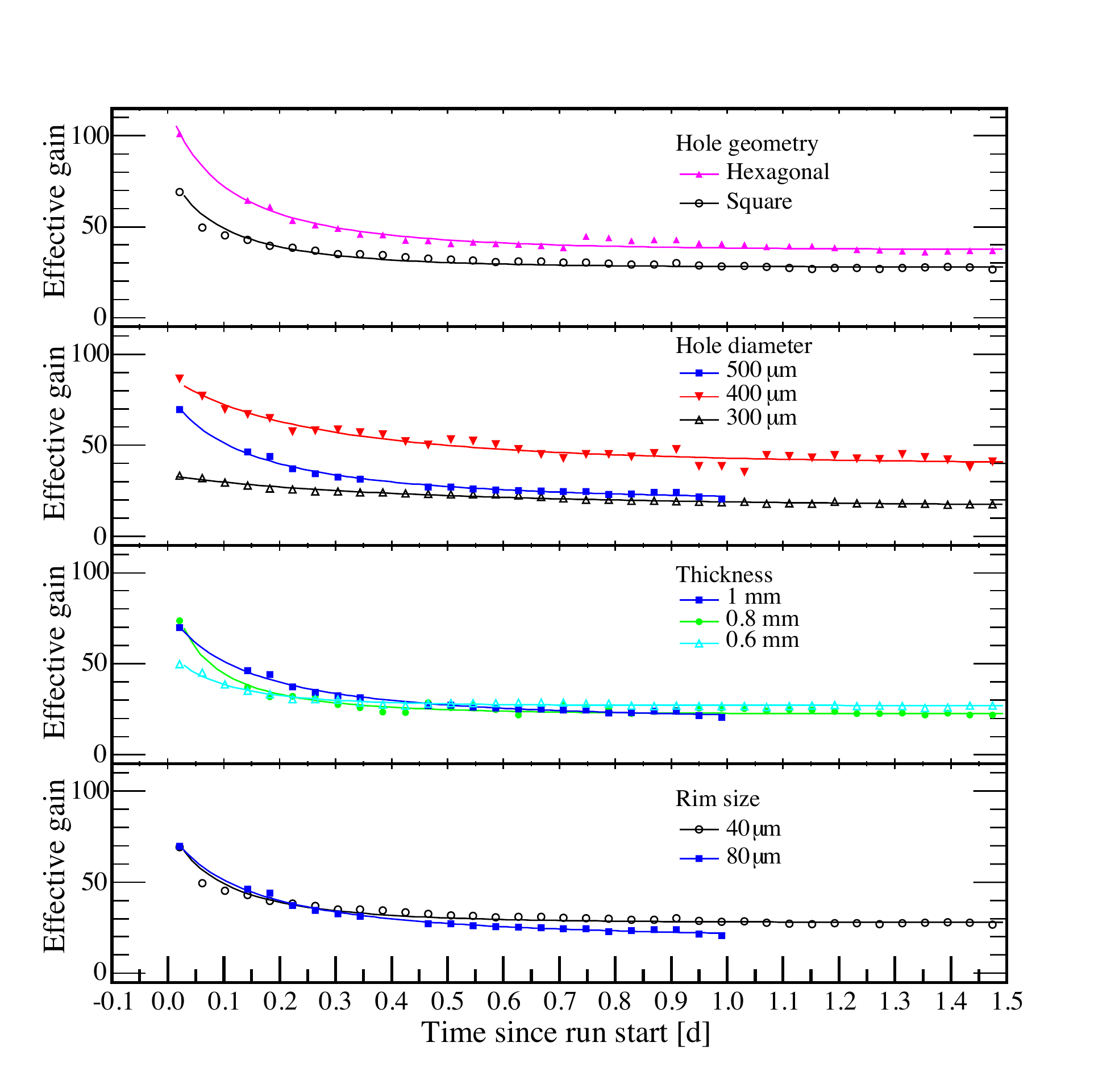} 

     \caption{Comparison of the time evolution of the effective gains
       for various LEMs in pure argon vapor at 87 K. The data points
       are fitted with the function described in the text.}
     \label{fig:gain_stability}
    \end{figure}

Immediately after finishing the LEM scans, the chamber was operated at
fixed electric field settings for at least one day to study the gain
evolution over time. The LEM fields were polarised according to the
values from \tabref{tab:stab_comp}.  For all LEMs, their initial gain
at $t_0$ were found to decrease and then stabilise after a typical
time of half a day as shown in \figref{fig:gain_stability}.  The
effective gain at $t_0$, defined as $G^{0}_{eff}$, is from the data
taken during the LEM electric field scan at the corresponding electric
field. All the data points are corrected for pressure variations and
normalised to a fixed pressure of 980~mbar in order to have a direct
comparison. We observe that the effective gain $G_{eff}(t)$ relaxes
from an initial gain of $G^{0}_{eff}$ to a stable value,
$G^{\infty}_{eff}$, after a characteristic time $\tau$. The time
evolution of $G_{eff}(t)$ can be empirically described as:
\begin{equation}\label{eq:time_func}
 G_{eff}(t)=\frac{G^{\infty}_{eff}}{1-(1-\frac{G^{\infty}_{eff}}{G^{0}_{eff}})\times e^{-t/\tau}}
\end{equation}

    This phenomenon was already observed and described in one of our
    previous publications \cite{Cantini:2013yba} and is understood to
    be a consequence of the charge accumulation on the insulator
    inside and around the hole.  A simulation of this effect along
    with the impact on the gain is given in
    Ref.~\cite{chargingup_sim}. During these long term operations,
    occasional discharges occurred between the top and bottom
    electrodes of the LEM. These discharges do not affect the
    evolution of the overall gain of the LEM. 
    
    \tabref{tab:stab_comp} summarises
    the total data taking and the number of discharges recorded by the
    power supply for all the LEMs.  As none or very few discharges
    occurred during typical run times of several tens of hours, we
    regard the LEM as a very stable apparatus operating at a stable
    effective gain of around 20. For the purpose of these measurements
    data taking was stopped after periods of a few days. We have
    already shown in \cite{Cantini:2013yba} that, once charged up, the
    LEM can be operated at the stable gain, $G^{\infty}_{eff}$, with
    very few discharges for periods of at least one month. The gain
    remains stable unless the run is stopped and the chamber is
    exposed to air which neutralises the charges stuck to the
    dielectric of the LEM.

\begin{figure}[h!]
     \centering
     \includegraphics[width=\textwidth]{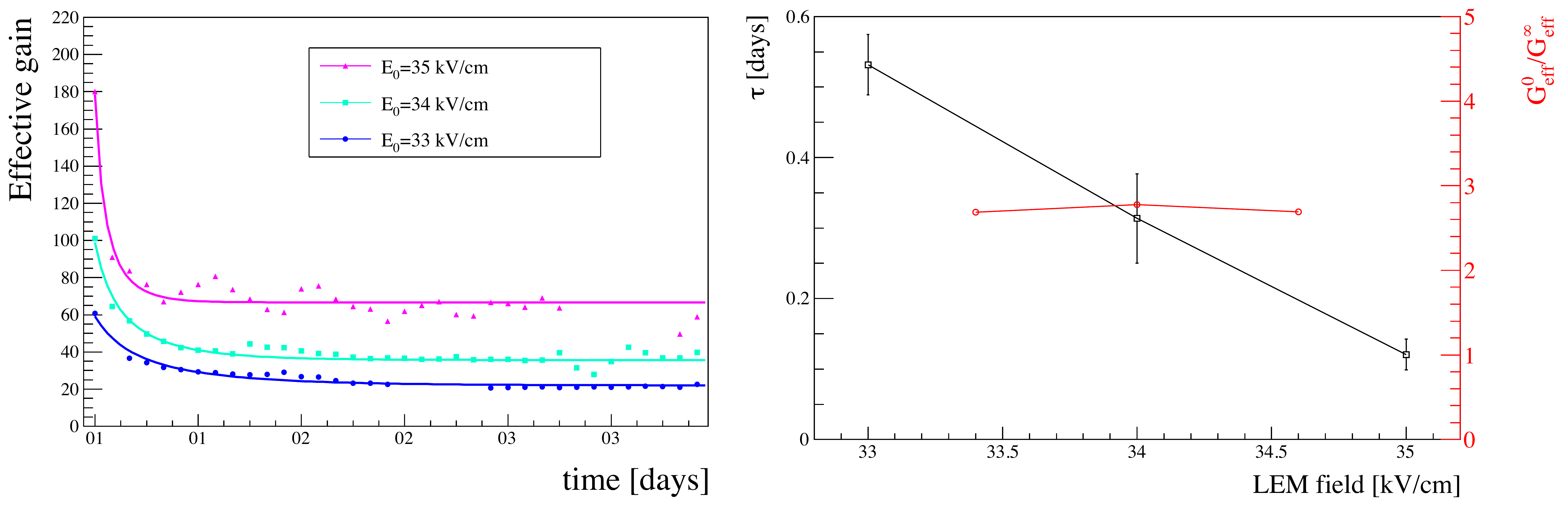}
     \caption{Left: evolution of the effective gain for the same LEM
       (LEM 3) operated in three independent runs with different
       electric field settings. Right: value of $\tau$ and the ratio
       of the initial to stable gain as a function of the applied
       electric field. }
     \label{fig:lem_stab_efield}
   \end{figure}

Since we compare LEMs operated at different effective gains we also
studied the response of the same LEM (LEM 3) operated at three
different electric fields in three independent runs. The results are
shown in \figref{fig:lem_stab_efield}. From these measurements we
conclude that 1) if the initial gain or applied electric field is
larger the LEM will charge up faster and 2) the ratio between the
initial gain and final gain is independent of the field at which the
LEM is operated.

   The impact from the different LEM parameters on the charging-up
   time $\tau$ as well as the ratio between initial and final gain
   have been investigated and are summarised in
   \tabref{tab:stab_comp}. They are discussed separately in the
   following paragraphs:
   
   \begin{itemize}
   \item[(i)] {\it Impact of the hole geometrical layout}: both LEMs
     were operated at the same $E_{0}$ but, because of their different
     electrical transparencies (see \secref{sec:lem_scan}), the LEM
     with the hexagonal hole layout has a much larger initial gain.
     Since we observe identical charging-up time constants we remark
     that the evolution of $\tau$ observed in
     \figref{fig:lem_stab_efield} may be an effect of the applied
     electric field rather than the initial effective gain. We also
     notice that the hole layout has no impact on the ratio between
     initial and final gain.

   \item[(ii)] {\it Impact of the hole diameter}: although operated at
     different electric fields the runs with LEMs of three different
     hole diameters have a similar stabilisation time around 0.5
     days. The ratio between the initial and final gain tend to
     decreases as the hole diameter becomes smaller. More discharges
     were observed during the runs with LEMs of smaller diameter holes
     compared to any other runs. This is especially true for the 300
     $\mu$m hole LEM which was operated at a relatively low gain for a
     duration which is comparable to other runs. One possible
     explanation is that smaller holes are more difficult to
     clean. Another explanation is that, in smaller diameter holes,
     the electrons from the avalanche are more susceptible to reach
     the edge of the hole where the electric field is larger and thus
     produce a discharge (see \secref{sec:lem_intro}).

   \item[(iii)]{\it Impact of thickness}: a reduction of the LEM
     thickness clearly results in a smaller ratio between the initial
     and final gain and a faster stabilisation time. As indicated in
     \figref{fig:lem_stab_efield} the faster charging-up time could be
     an effect of the larger electric field needed to polarise thinner
     LEMs. The reduction of the ratios,
     $\frac{G_{eff}^{0}}{G_{eff}^{\infty}}$, are evidence that the
     gain of the LEM evolves to a stable value because of free charges
     sticking inside the holes. As observed in \secref{sec:lem_scan},
     thinner LEMs cannot reach very high gains because of their
     shorter amplification lengths. On the other hand they have the
     advantage of relaxing faster and stabilising at relatively larger
     gains.

   \item[(iv)] {\it Impact of the rim size}: both LEMs were operated
     at similar initial gains and we clearly observe that the LEM with
     the smaller rim has a faster stabilisation time and a larger
     ratio between initial and final gain. Studies at different gas
     and temperature conditions
     ~\cite{Breskin2009107,Rimsize_Alexeev} showed similar
     conclusions: wider rim result in longer stabilisation time and
     larger ratios between initial and final gain. One possible
     explanation is that the exposed di-electric material increases
     with the rim size providing a larger area around the hole for
     free charges to stick.   
   \end{itemize}

\begin{table}[htb]
\renewcommand{\arraystretch}{1.2}
\begin{center}
\small
\begin{tabular}{llcp{1cm}p{1.4cm}p{1.5cm}p{1.6cm}p{1.0cm}p{0.7cm}ccc}
  \toprule 
  Parameter& Value &LEM & $E_{0}$ \newline{}   [kV/cm]& Run-time [hrs] & No. of \newline{} discharges & $\tau$ \newline{} [days]&$G_{eff}^{0}$ & $G _{eff}^{\infty}$&$\frac{G_{eff}^{0}}{G_{eff}^{\infty}}$\\
  \midrule  
  geometry&hexagonal&3&34 &110&0&0.32$\pm$0.07&99&35&2.7\\
  &square &5&34&52&0&0.30$\pm$0.02&65&27&2.4\\
  \midrule  
  hole&500 $\mu$m &2&38 &24&0 &0.53$\pm$0.05&70&20&3.5\\
  &400 $\mu$m &4&37&50&2&0.53$\pm$0.07&84&40&2.1\\
  &300 $\mu$m &6&33.5 &75&3&0.75$\pm$0.04&32&16&2.0\\
  \midrule  
  thickness&1 mm & 2&38 &24&0&0.53$\pm$0.05&70&20&3.5\\
  &0.8 mm & 1 &42 &82&0&0.24$\pm$0.02&73&22&3.3\\
  &0.6 mm & 7 &46 &95&1&0.18$\pm$0.01&51&27&1.9\\
  \midrule  
 rim size&80 $\mu$m &2&38&24&0&0.53$\pm$0.05&70&20&3.5\\
 &40 $\mu$m &5&34&52&0&0.29$\pm$0.02&65&27&2.4\\


  \bottomrule 
\end{tabular}
\end{center}
\caption{\label{tab:stab_comp} Summary of the results obtained from
  the longer term stability runs.The definitions
  of the variables are given in the text.}
\end{table}

\subsection{Uniformity of the gain}

\begin{figure}[htb]
  \centering
     \includegraphics[width=.7\textwidth]{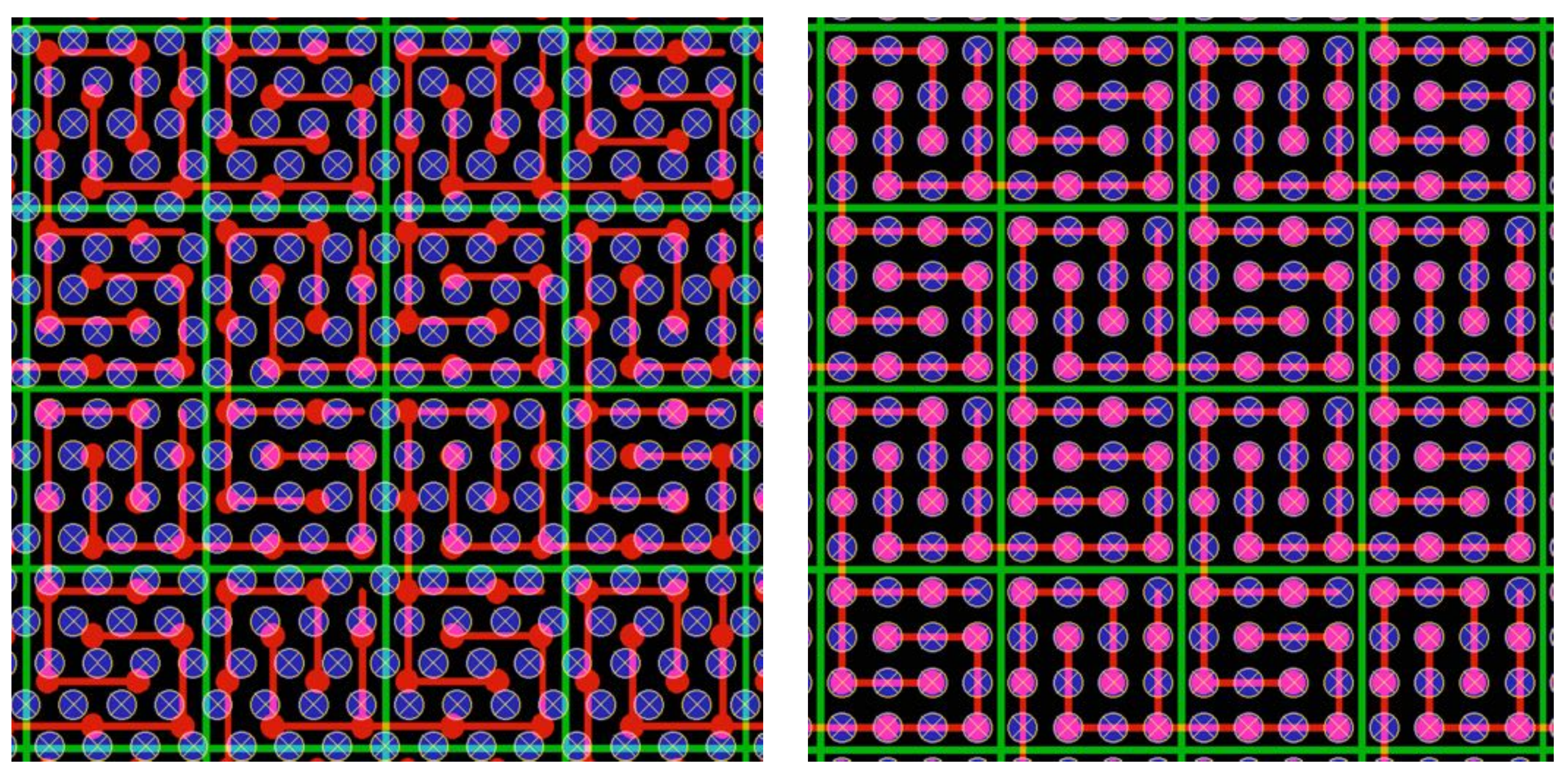}  
     \caption{Schematic of the matching between the extraction grid
       (green lines) the LEM holes (in opaque blue) and the multilayer
       PCB anode copper tracks (red pattern). In the case of the
       square LEM hole arrangement (right) the holes are aligned to
       the anode copper tracks.}
     \label{fig:lem_grid} 
   \end{figure}

In addition to the effective gain, we also studied if the hole layout
impacts the uniformity of the charge sharing between both
views. \figref{fig:lem_grid} shows the alignment between the
extraction grid, the LEM holes and the anode strips. The LEM square
and hexagonal hole layouts are compared: on the former the holes are
aligned with the pads of the anode readout strips whereas in the
hexagonal layout there is no specific matching between the holes and
the anode pattern. In both cases the extraction grid wires are
precisely aligned between the anode readout strips.
   Since the LEM with the hexagonal hole layout is the most
   transparent to drifting electrons and offers some of the best
   performances in terms of maximal achievable gain it is important to
   check that the hole layout does not degrade the imaging
   capabilities of the chamber. This could occur either by introducing
   large channel-to-channel fluctuations of the gain or compromising
   the symmetry of the charge sharing between both views.

\begin{figure}[h!]
     \centering
     \includegraphics[width=\textwidth]{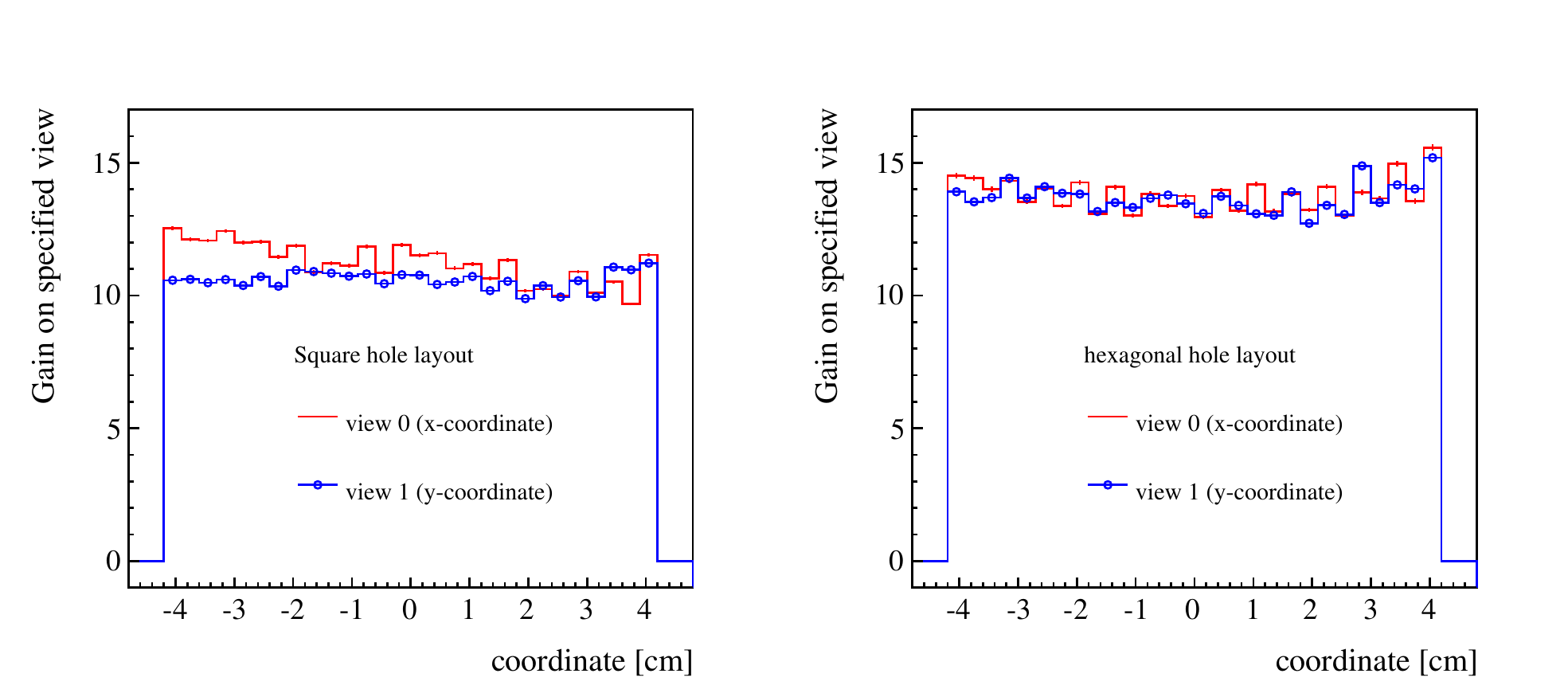} 
     \caption{Projections of the measured gains along both views in
       the case of the LEM with the square (left) and hexagonal
       (right) hole layout)}
     \label{fig:lem_unif}
   \end{figure}

   In Figure ~\ref{fig:lem_unif} we compare the gain measured on each
   strip along the $x$ and $y$ coordinates (or equivalently view 0 and
   view 1) for both hole layouts.
   In both cases, a strip-to-strip fluctuation is seen which is partly
   due to the limited precision on the alignment between the
   extraction grid and anode strips. The slight smooth trend on top of
   the strip-to-strip fluctuations is a consequence of variations in
   the LEM thickness. Even if, for future operations of larger
   detectors, this effect of the LEM thickness can be calibrated it
   will be important to precisely measure the thickness of the LEM
   insulator.

   In both cases, a deviation of within $\pm$ 10\% from the mean is
   achieved for all the strips in both coordinates. Taking into
   account also the variation in the electronics sensitivity which are
   typically of $\pm$ 5\% from channel to channel, the $\pm$ 10\%
   variation is within an acceptable region. For the square hole
   layout, we observed a 5\% difference in the charge collected by two
   views. This is due to the limited precision to match the anode pads
   with the LEM holes. For the hexagonal hole layout, the charge
   sharing between two views is within 1\% due to the random matching
   between the anode and LEM hole which smears out the difference. We
   can therefore conclude that no bias is introduced by using LEMs
   with hexagonal hole layouts.

\section{Conclusion}\label{sec:conclusions}
We have compared the response of LEMs with different characteristics
in pure argon vapor at 87 K.  In seven separate runs on a LAr LEM TPC
of $10\times 10$ cm$^2$ active area we have checked the impact of the
rim size, hole diameter, insulator thickness and hole arrangement on
both the maximal initial gain and its long term evolution. Each LEM
has a different response; the maximal achievable gain was 180
corresponding to $S/N\approx 800$ for MIPs over a 3~mm readout pitch. This impressive $S/N$ was
obtained with a LEM of 1 mm thickness, 40~$\mu$m rims and 500 $\mu$m
diameter holes arranged in a hexagonal pattern. With the chamber
exposed to cosmic rays, we observed that all the LEMs charge up in a
relatively short period of less than a day. Once charged up, the LEMs
can be operated at a stable gain of at least $\sim$20 with none or
very few discharges occurring across the electrodes.  We have also
shown that a uniform gain and excellent charge sharing can be achieved
in our setup. 

Our tests demonstrate that the LEM is a robust
apparatus that is well-understood and well-suited for operation in ultra-pure cryogenic
environments and that can maintain stable gains that match the goals of
future large-scale liquid argon TPCs. In this context we are now in
the process of manufacturing and testing LEMs and anodes with an
active area of 50$\times$50 cm$^2$ which are soon to be operated on
LAr LEM TPCs of much larger dimensions~\cite{Agostino:2014qoa}.

\end{document}